\documentclass[journal]{IEEEtran}
\usepackage[T1]{fontenc}
\usepackage{graphicx}
\usepackage{amsmath,amssymb,amsfonts}
\usepackage{algorithmic}
\usepackage{graphicx}
\usepackage{soul}
\usepackage{textcomp}
\usepackage{xcolor}
\usepackage{array}
\usepackage{multirow,multicol, makecell, booktabs}
\usepackage{amsmath}
\usepackage{comment}

\ifCLASSINFOpdf

\else

\fi

\begin{document}
\title{AsQM: Audio streaming Quality Metric based on Network Impairments and User Preferences}

\author{Marcelo~Rodrigo~dos~Santos,
        Andreza~Patrícia~Batista,
        Renata~Lopes~Rosa,
        Muhammad~Saadi,
        Dick~Carrillo~Melgarejo,~\IEEEmembership{Member,~IEEE}
        and~Demóstenes~Zegarra~Rodríguez,~\IEEEmembership{Senior~Member,~IEEE}%
\thanks{``This work was supported by the Brazilian National Council for Scientific and Technological Development (CNPq)''}
\thanks{M. R. dos Santos, A. P. Batista, R. L. Rosa and D. Z. Rodriguez are with the Federal University of Lavras, Lavras, MG, CEP 37202-618, Brazil (e-mail: marcelo.santos5@estudante.ufla.br, andreza.batista@estudante.ufla.br, renata.rosa@ufla.br, demostenes.zegarra@ufla.br).}
\thanks{M. Saadi is with University of Central Punjab, 54000, Lahore, Pakistan (e-mail: muhammad.saadi@ucp.edu.pk ).}
\thanks{D. C. Melgarejo is with School of Energy Systems, Lappeenranta-Lahti University of Technology (LUT), Lappeenranta, Finland (e-mail: dick.carrillo.melgarejo@lut.fi).}
}

\markboth{}%
{Shell \MakeLowercase{\textit{et al.}}: Bare Demo of IEEEtran.cls for IEEE Communications Society Journals}

\maketitle

\begin{abstract}
There are many users of audio streaming services because of the proliferation of cloud-based audio streaming services for different content. The complex networks that support these services do not always guarantee an acceptable quality on the end-user side. In this paper, the impact of temporal interruptions on the reproduction of audio streaming and the user’s preference in relation to audio contents are studied. In order to determine the key parameters in the audio streaming service, subjective tests were conducted, and their results show that user’s Quality-of-Experience (QoE) is highly correlated with the following application parameters, the number of temporal interruptions or stalls, its frequency and length, and the temporal location in which they occur. However, most important, experimental results demonstrated that users’ preference for audio content plays an important role in users’ QoE. Thus, a Preference Factor (PF) function is defined and considered in the formulation of the proposed metric named Audio streaming Quality Metric (AsQM). Considering that multimedia service providers are based on web servers, a framework to obtain user information is proposed. Furthermore, results show that the AsQM implemented in the audio player of an end user’s device presents a low impact on energy, processing and memory consumption.

\end{abstract}


%
\IEEEpeerreviewmaketitle

\section{Introduction}
\label{sec:introduction}
\IEEEPARstart{T}{he} presence of multimedia content in the internet has never been greater than today \cite{JAALAMA2021103996}. Most of the broadcast technologies and communication markets have turned to the emergent phenomena of Internet based solutions, such as social networks, online radio, music streaming and content sharing platforms \cite{9022990,9777989}. In every single minute, more than 300 hours of both audios and videos are uploaded to Internet to be distributed using different services; as a consequence, people are consuming more hours of these multimedia signals each year \cite{3210424.3210431}.

In recent years, web-based music streaming providers have increased and such fact has put the music streaming in the list of the contents most accessed by users via the Internet \cite{ishizaki2015understanding}. In general, it is expected that audio streaming service, only considering the cellular traffic, will reach 1.78 Exabyte per month by the end of 2025 \cite{9166225}, representing approximately 6.5$\%$  of the total amount of mobile data traffic. In many streaming services, the content is transmitted via broadcast worldwide, but using a connection for each user, in which packets must be sent to one listener at a time. Thus, network traffic increases based on the number of users; therefore, the probability of packet losses is higher \cite{SEYFOLLAHI2020107368}.

The occurrence of packet losses on transmissions over the User Datagram Protocol (UDP) \cite{protocol1980rfc} affects directly the quality of the content. UDP is used in real-time services, such as VoIP, which impairment characteristics were well studied in the past two decades \cite{rix2001perceptual}, and different quality evaluation methods were proposed \cite{chaudhery2017performance}. Nowadays, most streaming applications run over the Hypertext Transmission Protocol (HTTP) over Transmission Control Protocol (TCP),  and the multimedia quality metrics started to consider the peculiarities of TCP and its impairment characteristics to determine an accurate evaluation \cite{duanmu2018quality}.

Audio quality evaluation methods can be classified in two main categories, subjective methods that are based on the user’s evaluation of the content, and objective methods that are based on algorithms to estimate the signal quality \cite{8743302,8657699,8926509, 9281368,7797329}. Objective methods can also be classified according to their input types on speech-based, parametric-based \cite{9351640} and hybrid \cite{Rao2019}. The first one is sub classified in intrusive or nonintrusive methods \cite{8937202}. The intrusive method uses two audio signals, a reference and an impairment signal \cite{GOOSSENS2017109}. A nonintrusive method uses a single signal and it is the most appropriated for real-time services \cite{SONI202127}. In the later decade, some studies focused on video streaming services stated that temporal interruptions \cite{app10051793}, most known as stalling or pauses, is the major user’s Quality-of-Experience degradation factor \cite{rodriguez2014impact,9388867,9075375,rodriguez2016video,affonso2018speech}. Others well-known impairment factors are the initial delay \cite{rodriguez2012quality} and switching resolution events \cite{rodriguez2016video}. As shown in works \cite{rodriguez2012quality,duanmu2016quality,bampis2017study}, the temporal location in which the stalls occur also has a relevant impact on the users perceived quality. That means, stalls on the beginning of the content reproduction may have a different impact than stalls on other temporal segments. Therefore, an audio quality metric should assess not only the occurrence of stalls, but its temporal location in the audio. Also, ITU-T Rec. P.1203 \cite{robitza2018http} introduced a parametric bitstream-based quality assessment model \cite{8436879,9234526,7965631} for audiovisual streaming services over reliable transport; specifically, ITU-T Rec. P.1203.2 \cite{guzman2019automatic} presents an audio quality estimation module focusing on codec impairments as well as Internet protocol (IP) network impairments \cite{8475048}. This audio module predicts mean opinion scores (MOS) based on a 5-point absolute category rating (ACR) scale \cite{itu1999subjective}.

On another hand, some studies in multimedia content pointed that user’s QoE does not only depend on technical aspects, user subjectivity, such as preferences on multimedia content \cite{jaramillo2016content,vandenbroucke2015subjective,laiche2021machine} play an important role in the user experience prediction. In addition, It is important that in subjective tests, the assessors’ profiles information, such as: age, gender, expectations, emotions and preferences \cite{rosa2013sentimeter,rosa2013sentimeter2,rodriguez2014video} need to be considered. In this sense, the user preference for audio content is taken into account in this work, in order to improve an objective audio quality metric.

In this context, the main contribution of this paper is to propose a new audio quality metric specifically designed to address the impairments of audio streaming over TCP/IP, named Audio streaming Quality Metric (AsQM). The proposed metric considers the following criteria: (a) number of stalls, (b) stalls duration, (c) temporal location of the stalls, (d) initial buffering duration, and (e) user preference on audio content. The latest one is proposed as an adjustment factor, let AsQM works in case user preference information is not available, and also it can be used by other audio quality metrics. The performance assessment of our proposal is performed using subjective tests, and also compared with the results obtained by the method described in ITU-T Rec. 1201.2 \cite{garcia2013parametric}. Moreover, the proposed audio quality metric was implemented in a handheld mobile electronic device as an application, to evaluate its performance, and the experimental results showed that the AsQM consumes negligible resources from the mobile device. Finally, in order to extract user audio content preference information, an architecture of audio streaming service is introduced.

The rest of this manuscript is structured as follows. Section II presents a review of Audio Quality Assessment Methods. Section III introduces the proposed Audio Quality Model used to determine the AsQM. Section IV shows the implementation of the tests. Experimental results are showed in Section V. Finally, the conclusions are described in Section VI.

\section{The Proposed Audio streaming Quality Metric}
 
\begin{figure}[ht]
\centering
       \includegraphics[width=0.99\linewidth]{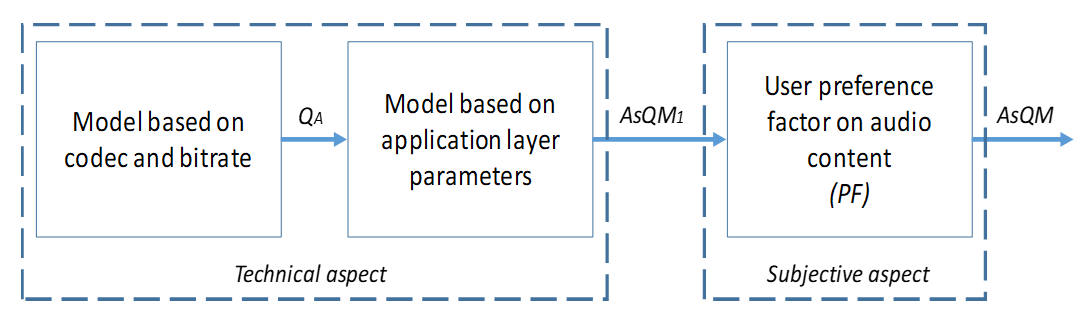}
\caption{Block diagram of the proposed $AsQM$.}
\label{fig:fig1}
\end{figure}

The main components of the proposed $AsQM$ are introduced in Fig. \ref{fig:fig1}.

\vspace{\baselineskip}
\vspace{\baselineskip}

As can be observed, the $AsQM$ is composed by two main blocks depicted in dash-lines. The first one corresponds to the module that considers technical aspects, such as codec characteristics and application layer parameters to estimate a quality index, named $AsQM\textsubscript{1}$. The quantification of codec degradation is based on the specifications of ITU-T Rec P.1203.2, and the application layer parameters considers the effect of stalls and their temporal locations. The second module is related to the user subjectivity, and it works as an adjustment factor that takes the user preference on audio content into account. It is important to note that the $AsQM$ formulation in two stages, permits the second module can be used for another audio quality metric, in case user preference is known.

The proposed $AsQM$ is determined according to the following relation

\begin{equation}
   AsQM = AsQM_{1} \times PF
   \label{eq:eq1}
\end{equation}
where $PF$ represents a function regarding the user preference factor, and $AsQM\textsubscript{1}$ is given by

\begin{equation}
 AsQM_{1} = Q_A - I_D - I_S 
 \label{eq:eq2}
\end{equation}
where $Q_A$ represents the audio quality reached by a specific codec at a certain bitrate and considering a 5-point quality scale, $I_D$ represents the impairment factor regarding the initial delay, and $I_S$ is a relation to calculate the impairment factor due to stalls happened during the audio streaming.
  
To determine each component of $AsQM$ formulation, with the exception of $Q_A$, subjective tests on audio quality were carried out in this work. For a better explanation of the methodology followed, we divided the test procedure in phases I, II and III, which are described as follow.
  
\subsection{Phase I: Determination of audio codec impairment}
\addcontentsline{toc}{subsection}{Phase I: Determination of audio codec impairment}
   
As stated before, for the determination of audio codec impairment, the following relations presented in ITU-T Rec P.1203.2 are used in this work
  
\begin{equation}
  Q_A = MOSfromR (100 - QcodA)
  \label{eq:eq3}
\end{equation}
where $Q_A$ represents the audio quality of the codec used in the streaming without considering any degradation. $MOSfromR(X)$ is an operator to transform R-scale values to 5-point MOS scale, and $QcodA$ represents the actual impairment of audio codec. 

The $QcodA$ relation is presented as follow:
 
\begin{equation}
QcodA = \alpha\textsubscript{1} \times exp(\alpha\textsubscript{2} \times BR) + \alpha\textsubscript{3}
\label{eq:eq4}
\end{equation}  
where $\alpha\textsubscript{1}$, $\alpha\textsubscript{2}$ and $\alpha\textsubscript{3}$ are coefficients of the model for a specific audio codec, and $BR$ is the audio bit rate expressed in kbps.

The $MOSfromR(X)$ is defined in (\ref{eq:eq5}).

\begin{equation}
    \begin{split}
        & MOSfromR(X) = M_{MIN}+(M_{MAX}-M_{MIN})X\div100\\ 
        & X (X -60)(100 - X) 7.10^{-6}
        \label{eq:eq5}
    \end{split}
\end{equation}
where $X$ represents an R-scale quality score; and $M_{MIN}$ e $M_{MAX}$ are the minimum and maximum MOS index values permitted, they are equal to 1.05 and 4.9, respectively.

In our experimental tests, two audio codecs are used, AAC-LC and HE-AAC-v2, which coefficients introduced in (4) and bit rates supported by each one of them are presented in Table \ref{eq:eq1}.
 
\vspace{\baselineskip}
\begin{table}[ht]
\caption{COEFFICIENT VALUES USED TO ESTIMATE THE AUDIO CODEC IMPAIRMENT}
\label{tab:tab1}
\centering
\resizebox{\columnwidth}{!}{%
\begin{tabular}{>{\centering\arraybackslash}m{2.05cm}>{\centering\arraybackslash}m{1.05cm}
                >{\centering\arraybackslash}m{1.05cm}>{\centering\arraybackslash}m{1.05cm}>{\centering\arraybackslash}m{1.45cm}}
    \hline
    \textbf{Audio Codec} & \textbf{\large{$\alpha\textsubscript{1}$}} & \textbf{\large{$\alpha\textsubscript{2}$}} & \textbf{\large{$\alpha\textsubscript{3}$}} & \textbf{Bit rate - BR (kbps)}\\
    \hline
    \hline
    {AAC-LC} & {100} & {-0.05} & {14.6} & {32-576}\\
    {HE-AAC-v2} & {100} & {-0.11} & {20.06} & {16-96}\\
    \hline
\end{tabular}
}
\end{table}
\vspace{\baselineskip}

Then, the $Q_{A}$ values can be calculated for both codecs using (3), (4) and (5).

\subsection{Phase II: Study of the impact of initial delay on the global perceived quality}
\addcontentsline{toc}{subsection}{Phase II: Study of the impact of initial delay on the global perceived quality}
   
\begin{figure}[ht]
	\begin{center}
		\includegraphics[width=3.5in,height=1.97in]{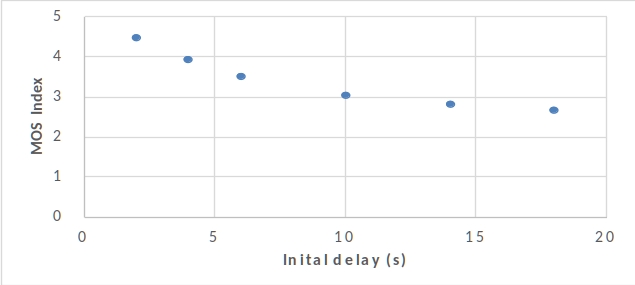}
		\caption{Initial buffering delay and its effects on users’ QoE.}
		\label{fig:fig2}
	\end{center}
\end{figure}

The impact of the initial delay on the global user’s perceived audio quality was analyzed. Preliminary subjective tests that considered different initial delay lengths were performed. A 1-minute audio length with 6 different initial delay lengths was used as test material, which results are presented in \ref{fig:fig2}. In order to minimize the codec impairments, the audio codec used was the AAC-LC at its maximum bit rate.

Fig. \ref{fig:fig2} shows that the relation between the MOS quality index and the initial delay length can be approximated to an exponential function. The impairment caused by the initial delay can be modeled as a logarithmic function.

However, the degradation caused by the initial delay can depend on the audio length; therefore, the ratio between initial delay and total video length should be considered as presented in (\ref{eq:eq6}).

\begin{equation}\label{eq:eq6}
I_{D}=-k\cdot\ln \left(\frac{c\cdot D_{L}}{T_{L}}\right)
\end{equation}
where $I_D$ is the impairment added by the initial delay; k is a constant for scaling purposes; $D_L$ represents the initial delay length in seconds; $c$ indicates an exponential decaying factor; and $T_L$ represents the audio total length in seconds. It is important to note that $I_D$ is determined from subjective test results as the difference between $Q_A$ and $MOS$ index.
  
\subsection{Phase III: Study of the relation between pauses in audio streaming and user’s QoE}
\addcontentsline{toc}{subsection}{Phase III: Study of the relation between pauses in audio streaming and user’s QoE}
   
This work aims to study the influence of stalls during an audio transmission. The number of pauses and their temporal location influence the user’s QoE in audio services \cite{affonso2018speech,9123118}.
  
Streaming applications may implement different strategies to minimize the impact of network losses; then, it is very important to understand which kinds of scenarios have greater impact on the user’s perceived quality. Hence, it was necessary to build several scenarios or impairment audios containing stalls at different temporal location and with certain duration as presented in the previous section.

The proposed $I_S$ model considers the following parameters, the initial buffering, the number of stalls, the stall lengths, and the impairment weight of the temporal segment in which the stalls occur. The last one is not considered into the model given by ITU-T Rec. 1201.2 \cite{garcia2013parametric}.

\begin{figure}[ht]
	\begin{center}
		\includegraphics[width=1.00\linewidth]{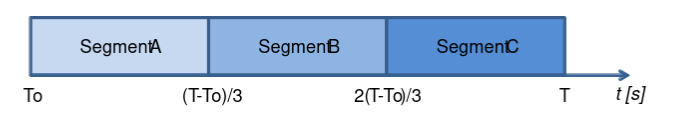}
		\caption{Definition of temporal segments: Initial audio segment (Segment A); the intermediate audio segment (Segment B); and the final audio segment (Segment C).}
		\label{fig:fig3}
	\end{center}
\end{figure}

For this proposal, to determine the temporal locations of the stalls, three temporal segments were defined: (a) segment A, which represents the initial audio segment; (b) segment B, the intermediate segment; and (c) segment C, which represents the final audio segment. These segments are illustrated in Fig.~\ref{fig:fig3}, in which \textit{T\textsubscript{0}} represents the instant in which the audio player starts after of the initial buffering period, and T\ corresponds to the instant in which the audio quality is evaluated.  Then, temporal segments are independent of the audio file length and each temporal segment length can be calculated at any instant. Also, the number of temporal segments influence the number of the total audio files to be created for testing, thus, it was restricted to three.

As previously mentioned, the goal of the definition of these temporal segments is to investigate the impact of stalls happened in each temporal segment on the global audio quality.

With this information, the \textit{I\textsubscript{S}} model that only depends on stall characteristics is modeled. Because there are many degradation factors, assessors of subjective tests are asked to evaluate the global quality. Additionally, for useful purposes, the 5-point scale of MOS was considered, which is one the most accepted quality scale used in voice quality assessment. For this, an exponential function was used, and it is presented in (\ref{eq:eq7}).

\begin{equation} \label{eq:eq7}
I_s = Q_A - c\cdot \exp{\left(\sum_{i=1}^{n}\frac{S_{i}\cdot L_{i}\cdot D_{i}}{T_{i}}\right)}  
\end{equation}
where, \textit{Q\textsubscript{A}} is the quality reached by a specific audio codec in ideal conditions, $c$ represents a constant, \textit{S\textsubscript{i}} represents the number of stalls; \textit{L\textsubscript{i}} is the average length of stalls, measured in seconds, which happens in the same temporal segment; \textit{D\textsubscript{i}} is a degradation degree that each temporal segment adds to the total audio degradation, it is used as a weigh factor; \textit{T\textsubscript{i}} is time period in seconds of each temporal segment; n is the number of temporal segments of an audio; in this work, three segments are considered for all the tests. It is worth noting that in each test scenario, $I_S$ is determined from subjective test results as the difference between $Q_A$ and \textit{MOS} index

The results of subjective audio tests determined the weight factor related to the degradation degree for each temporal segment $D_{A}$, $D_{B}$ and $D_{C}$.

In total, 53 different impairment stall models (or test scenarios) were implemented, and also two audio codec were used, which are explained in detail in the next section. An average $MOS$ index for the 3 assessed audios with the same impairment model (considering the 3 audio content categories) in the subjective test was used to calculate the parameters introduced in the \textit{I\textsubscript{S}} relation. For example, the following relation corresponds to the impairment model number 1 that has an average result called $I_{S-1}$:

\begin{equation}\label{eq:eq8}
    \begin{split}
    & L_n(Q_{A-1}-I_{S-1})=L_n(C)+\frac{S_A\cdot {L_A}\cdot {D_A}}{T_A}+\\
    & \frac{S_B\cdot{L_B}\cdot{D_B}}{T_B}+\frac{S_C\cdot{L_C}\cdot{D_C}}{T_C}
    \end{split}
\end{equation}
   
A linear system with unknown variables and 53 equations was obtained using (3). Later, the least squared method, specifically the pseudo-inverse, was used to resolve this equation system. \textit{$``$D\textsubscript{X}$"$}  represents the degradation weight of the temporal segment \textit{$``$X$"$}  to be determined. Note that the variables \textit{I\textsubscript{SX}}, \textit{S\textsubscript{X}}, \textit{L\textsubscript{X}} and \textit{T\textsubscript{X}} are known for each model. Also, it is important to stress that user preference was not considered.

An over determined equation linear system was obtained considering the 53 impairment models and (\ref{eq:eq8}), which is represented by:

\vspace{\baselineskip}
\begin{equation}
    \begin{bmatrix}
        1 & t_{1,2} & \cdots & t_{1,4}\\
        1 & t_{2,2} & \cdots & t_{2,4}\\
        \vdots & \vdots & \ddots & \vdots\\
        \vdots & \vdots  & \ddots & \vdots\\
        1 & t_{53,2} & \cdots & t_{53,4}
    \end{bmatrix}
    \times
    \begin{bmatrix}
        L_n(C)\\
        D_A\\
        D_B\\
        D_C
    \end{bmatrix}
    =
    \begin{bmatrix}
        L_n(Q_{A-1}-I_{S-1})\\
        L_n(Q_{A-2}-I_{S-2})\\
        \vdots\\
        L_n(Q_{A-53}-I_{S-53})
    \end{bmatrix}
\end{equation}
\vspace{\baselineskip}
   
The variables \textit{t\textsubscript{1,2}} to \textit{t\textsubscript{1,4}} presented in (4) represent the first impairment model; \textit{t\textsubscript{2,2}} to \textit{t\textsubscript{2,4}} represent the second model and so on. Solving this equation linear system, the values of c, \textit{D\textsubscript{A}}, \textit{D\textsubscript{B}} and \textit{D\textsubscript{C}} were obtained. 
  
\subsection{Phase IV: study of the relation between user preference on audio content and user’s QoE}
\addcontentsline{toc}{subsection}{Phase IV: study of the relation between user preference on audio content and user’s QoE}
   
\begin{figure}[ht]
	\begin{center}
		\includegraphics[width=3.5in,height=1.97in]{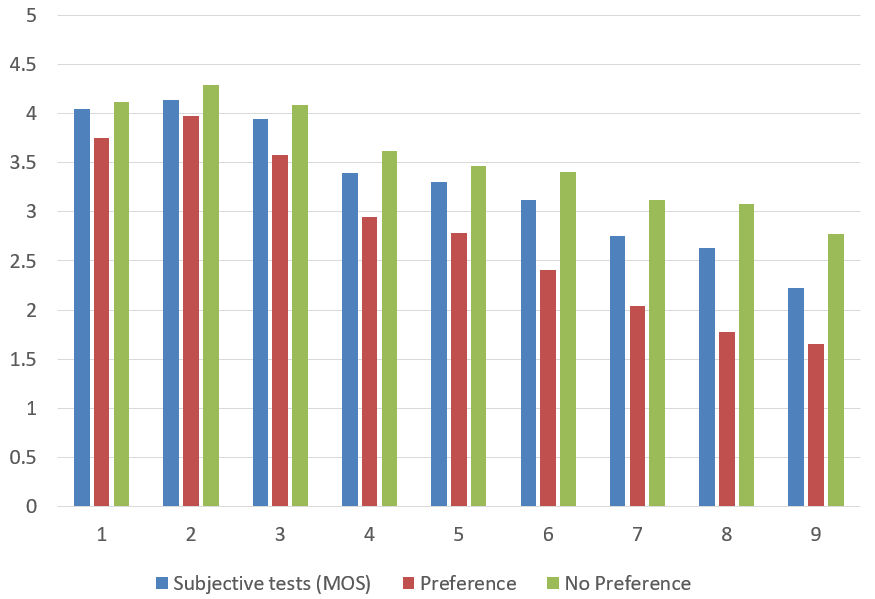}
		\caption{Preliminary subjective test results of nine audio sequences that permit to evaluate the impact of user preference on the audio quality index.}
		\label{fig:fig4}
	\end{center}
\end{figure}

In this phase, eight audios are used in the test. Each one had a duration of 120 seconds without considering pauses. The experimental results of the preliminary subjective tests are presented in Fig. \ref{fig:fig4}.

The MOS index presented in Fig. \ref{fig:fig4} shows that assessors with preference for a specific audio content type scored different values compared with assessors without preference for the same content type. The results demonstrate the importance of considering the user’s preference for audio content type in an objective audio quality metric. It is important to note that to the best of our knowledge, current objective audio quality metrics do not consider the user preference.

The \textit{Preference Factor (PF)} is conceptualized in this section. PF adjusts the audio quality score given by an algorithm. To reach a better relation to users' QoE, the algorithm that only considers codec and network parameters ($ASQM\textsubscript{1}$) is complemented with the proposed PF.

The subjective test results showed that \textit{PF} depends on the user preference, the audio content type, and also depends on the score level obtained by the objective audio quality metric. In this work, three content categories were considered, music, sport and news.
   
Twenty different audios for each category were evaluated. For each audio assessed in auditory test, a MOS score is obtained. These MOS results were used to establish the ratios between the MOS value given by both the users with and without preference. The results are analyzed considering each audio category used in the tests, and then compared with the results obtained by audio quality algorithms.

The variables $ARp_i$ and $ARnp_i$ represent the ratios between the MOS values scored by users with and without preference by the audio category, respectively. Both variables are introduced in (\ref{eq:eq10}) and (\ref{eq:eq11}), respectively.
  
\begin{equation}\label{eq:eq10}
    ARp_{i} = \frac{MOS_{i}^{preference}}{MOS_{i}^{mean}}
\end{equation}

\begin{equation}\label{eq:eq11}
    ARnp_{i} = \frac{MOS_{i}^{no-preference}}{MOS_{i}^{mean}}
\end{equation}
\vspace{\baselineskip}
   
The mean value represented by $MOS_{i}^{mean}$ considers the all users independent of their preferences for the audio test $``i"$. We considers that the maximum value for $``i"$  is 20 for each audio content category, because test development constraints. In the subjective tests, the same number of assessors with preference and no-preference was considered, in each audio assessed $``i"$ ; then the relation between  $ARp\textsubscript{i}$ and $ARnp\textsubscript{i}$ is formulated in (\ref{eq:eq12}).
  
\begin{equation}\label{eq:eq12}
    ARnp_{i} = 2 - ARp_{i}
\end{equation}
   
Let $PF_{p}^{CT}$ represent the $PF$ function based on the $ARp\textsubscript{i}$ values for each audio category. These values are obtained from subjective testes; thus, $PF_{p}^{CT}$ can be adjusted empirically by:

\begin{equation}\label{eq:eq13}
    PF_p^{CT}=\alpha{\cdot}\ln(MOS_{p}^{mean})+\beta
\end{equation} 
   
Where, $CT$ represents the audio content category and the $p$ index represents that user has preference for $CT$.  Table III presents the values for each variable used in (\ref{eq:eq13}).
  

\begin{table}[ht]
\caption{Variables values for the function considering different categories}
\label{tab:tab2}
\centering
\resizebox{\columnwidth}{!}{%
\begin{tabular}{>{\centering\arraybackslash}m{3.15cm}>{\centering\arraybackslash}m{2.15cm}>{\centering\arraybackslash}m{2.15cm}}
    \hline
    \textbf{AUDIO CATEGORY (CT)} & \textbf{\large{$\alpha$}} & \textbf{\large{$\beta$}}\\
    \hline
    \hline
    {Music} & {0,423} & {0.197}\\
    {Sport} & {0.699} & {0.428}\\
    {News} & {0.481} & {0.256}\\
    \hline
\end{tabular}
}
\end{table}
   
The maximum error obtained by using $\alpha$ and $\beta$ values in (13) for music, sport and news categories were 0.03, 0.03 and 0.04, respectively.
 
A function that represents the $ARnp\textsubscript{i}$ values called $PF_{np}^{CT}$ is determined by (\ref{eq:eq14}), in which the same $\alpha$ and $\beta$ values presented in Table II are considered.
 
\begin{equation}\label{eq:eq14}
    PF_{np}^{CT}=2-\alpha{\cdot}\ln(MOS_{p}^{mean})+\beta
\end{equation} 
   
The preference factor functions can be used in different audio streaming service implementations, in which the subjective  value is replaced in (\ref{eq:eq13}) and (\ref{eq:eq14}) for the MOS index obtained by an objective metric. In this work, that objective metric is represented by $AsQM\textsubscript{1}$, and using (\ref{eq:eq1}), the proposed $AsQM$ is determined.

It is important to note that to implement the $AsQM$ metric, the user’s preference needs to be stored in the audio server and each audio sample only belongs to a sole audio category. Also, it is important to note that the 5-point scale should adopted by the objective metric.
  
\section{Test Implementation}
\addcontentsline{toc}{section}{Test Implementation}
   
The audio database characteristics are described in this section. These audios are used as test material to perform the audio quality subjective tests. Later, the implementation of the audio player is treated.
  
\subsection{Audio Database Characteristics}
\addcontentsline{toc}{subsection}{Audio Database Characteristics}
   
Before to present the audio database used in the subjective tests, the impact of different network impairments on the audio streaming service is investigated; later, the criteria used to build the audio database is explained in detail.

\subsubsection{Network impairment and stall patterns}
\addcontentsline{toc}{subsubsection}{Network impairment and stall patterns}
   
Networks can suffer different impairments, in audio streaming services based on TCP those impairments are manifested as stalls during the reproduction; due of the network volatility, different stall patterns appear. In \cite{bampis2017study} are proposed some stall distributions, but they considered fixed stall lengths which is not a realistic impairment scenario.
  
In this research; firstly, a network scenario was implemented and the bandwidth and packet loss rate (PLR) parameters were used to create different stall patterns. Secondly, based on the network emulation results, the temporal location and the range of the stall lengths were defined. 
 
In order to implement different PLR distributions, the Gilbert-Elliot model was used as presented in (15) and (16):

\begin{equation}\label{eq:eq15}
    p=P(q_t=B\mid{q_{t-1}=G})
\end{equation}

\begin{equation}\label{eq:eq16}
    r=P(q_t=G\mid{q_{t-1}=B})
\end{equation}
where, is the probability to pass from a Bad state ($B$) that indicates packet loss to a Good state ($G$) that represents a success in the packet delivery; is the probability to pass from $G$ state to $B$ state; and represent the states at the instants \textit{t} and \textit{t-1}, respectively.

Thus, with the variation of and is possible to calculate the PLR, and also different packet losses distributions can be obtained as presented in (17):

\begin{equation}
    PLR = \frac{p}{p+r}
\end{equation}

\begin{figure}[ht]
	\begin{center}
		\includegraphics[width=2.96in,height=1.42in]{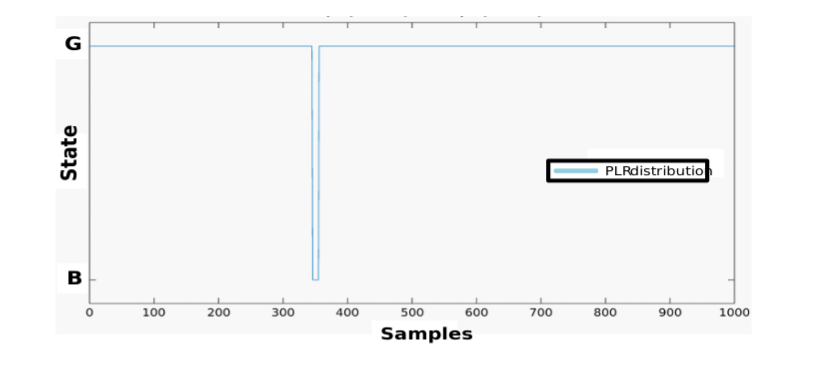}
		\caption{Packet loss distribution considering PLR=1$\%$ , p = 0.10101$\%$  and q = 10$\%$ .}
		\label{fig:fig5}
	\end{center}
\end{figure}

\begin{figure}[ht]
	\begin{center}
		\includegraphics[width=2.97in,height=1.54in]{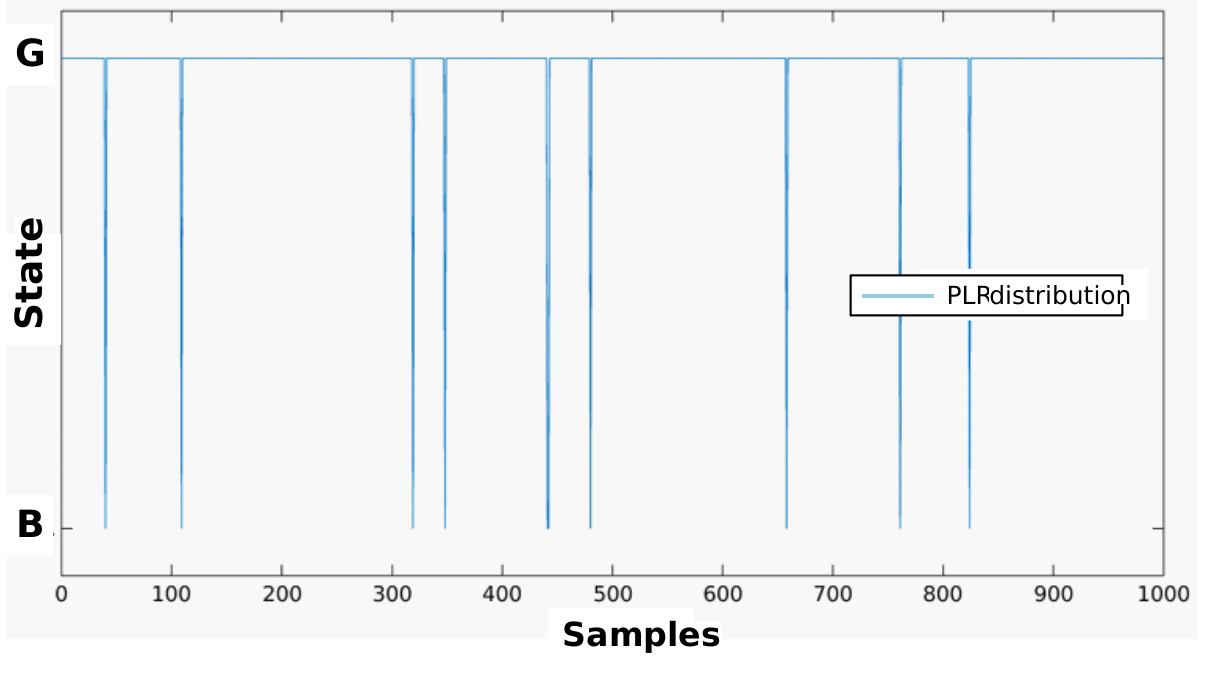}
		\caption{Packet loss distribution considering PLR=1$\%$ , p = 0.75758$\%$  and q = 75$\%$ .}
		\label{fig:fig6}
	\end{center}
\end{figure}

In Fig. \ref{fig:fig5}, a packet loss distribution is presented, considering a PLR of 1$\%$  and \textit{p}= 0.10101$\%$  and \textit{q} = 10$\%$. In Fig. \ref{fig:fig6} another packet loss distribution is showed that considers the same PLR of 1$\%$  but now with \textit{p}= 0.75758$\%$  and \textit{q} = 75$\%$ . In both figures the number of samples was limited to 1000.

As can be observed in Fig. \ref{fig:fig5} and Fig. \ref{fig:fig6}, depending on the probability of moving from a Good state to a Bad state and vice versa, the temporal distribution of packet losses varies. In audio streaming service, those different packet loss distributions affect in different manner; distributions that are similar to the presented in Fig. \ref{fig:fig5} cause longer stalls but a few number of them; conversely, similar distributions to the depicted in Fig. \ref{fig:fig6} originate a higher number of shorter stalls.
   
Additionally, the bandwidth capacity of the network transmission was changed, considering 200$\%$, 100$\%$, 90$\%$, 80$\%$, 70$\%$ and 50$\%$ of the minimum transmission rate of the corresponding audio file. These tests were mainly performed to create different scenarios to determine the initial buffering delay duration.

After the experimental tests, different stall patterns were obtained in the audio streaming service at the end-user device, and they were considered to build the audio DB for testing. 
  
\vspace{\baselineskip}
\subsubsection{Criteria to build the audio database}
\addcontentsline{toc}{subsubsection}{Criteria to build the audio database}
   
An audio database was built, considering the following criteria: number of stalls, temporal position of stalls, stall lengths and initial buffering delay. 
  
The number of stalls and their lengths were determined according to the network emulation results. It was observed that the minimum and maximum stall lengths were 1 sec. and 7 sec., respectively. Also, the minimum and maximum numbers of stall were 1 and 12 stalls. Additionally, the minimum and maximum initial buffering delays were 1 and 9 sec. Based on that information, different impairment patterns were defined. For a better understanding and data presentation, they were classified in different impairment level groups as presented in Table III.



\begin{table}[ht]
\caption{Description of the impairment groups based on the characteristics of stalls}
\label{tab:tab3}
\centering
\resizebox{\columnwidth}{!}{%
\begin{tabular}{>{\centering\arraybackslash}m{1.61cm}>{\centering\arraybackslash}m{2.15cm}
                >{\centering\arraybackslash}m{2.15cm}>{\centering\arraybackslash}m{2.15cm}}
    \hline
    \textbf{PARAMETER} & \textbf{LOW IMPAIRMENT/ (DESCRIPTION)} & \textbf{MEDIUM IMPAIRMENT/ (DESCRIPTION)} & \textbf{HIGH IMPAIRMENT/ (DESCRIPTION)}\\
    \hline
    \hline
    Initial delay [s] & 1 – 3 / (Id-L) & 4 – 6 / (Id-M) & 7 - 9 / (Id-H)\\
    Number of stalls & 1 – 4 / (Ns-L) & 5 – 8 / (Ns-M) & 9 – 12 / (Ns-H)\\
    Stall lengths [s] & 1 – 2 / (Sl-L) & 3 – 5 / (Sl-M) & 6 –  7 / (Sl-H)\\
    \hline
\end{tabular}
}
\end{table}
\vspace{\baselineskip}
   
The impact of the initial delay on the global audio quality was evaluated separately. For this, three audio lengths of 30, 60 and 90 seconds were evaluated and 3 initial delay lengths from each \textit{Id-L, Id-M e Id-H} were used, totalizing 27 audios with an initial buffering.

With the impairment level groups for the number and length of stalls parameters introduced in Table III, different impairment models were created. The\ characteristics of each impairment model or scenario are presented in Table IV.  

Additionally, three different audio content types were used: music, sport and news. These three audios do not contain any degradation type and they are named original audios. All the impairment models presented in Table IV were applied to each of the audio content type. As can be observed, the total number of impairment models is 53. Also, two different audio codecs were used; therefore, the total number of audios containing stalls that will be used as test material is 318.

It is worth noting that each audio contains a complete idea about some topic to avoid any assessors’ dissatisfaction, and all of them are in Portuguese language that is the native language of all assessors.

\begin{table}[ht]
\caption{DESCRIPTION OF THE IMPAIRMENT MODELS BASED ON THE CHARACTERISTICS OF STALLS}
\label{tab:tab4}
\centering
\resizebox{\columnwidth}{!}{%
    \begin{tabular}{>{\centering\arraybackslash}m{1.43cm}>{\centering\arraybackslash}m{0.80cm}>{\centering\arraybackslash}m{0.80cm}>{\centering\arraybackslash}m{0.80cm}>{\centering\arraybackslash}m{0.80cm}>{\centering\arraybackslash}m{0.80cm}>{\centering\arraybackslash}m{0.80cm}}
    \hline
    \multirow{2}{1.43cm}{\centering \textbf{Impairment Model}}
    & \multicolumn{2}{c}{\textbf{Segment A}} & \multicolumn{2}{c}{\textbf{Segment B}} & \multicolumn{2}{c}{\textbf{Segment C}}\\
    \cline{2-7}
    & Number & Length & Number & Length & Number & Length\\
    \hline
    \hline
    M1 & Ns-L & Sl-L & $-$ & $-$ & $-$ & $-$\\
    M2 & $-$ & $-$ & Ns-L & Sl-L & $-$ & $-$\\
    M3 & $-$ & $-$ & $-$ & $-$ & Ns-L & Sl-L\\
    M4 & Ns-L & Sl-M & $-$ & $-$ & $-$ & $-$\\
    M5 & $-$ & $-$ & Ns-L & Sl-M & $-$ & $-$\\
    M6 & $-$ & $-$ & $-$ & $-$ & Ns-L & Sl-M\\
    M7 & Ns-L & Sl-H & $-$ & $-$ & $-$ & $-$\\
    M8 & $-$ & $-$ & Ns-L & Sl-H & $-$ & $-$\\
    M9 & $-$ & $-$ & $-$ & $-$ & Ns-L & Sl-H\\
    M10 & Ns-M & Sl-L & $-$ & $-$ & $-$ & $-$\\
    M11 & $-$ & $-$ & Ns-M & Sl-L & $-$ & $-$\\
    M12 & $-$ & $-$ & $-$ & $-$ & Ns-M & Sl-L\\
    M13 & Ns-M & Sl-M & $-$ & $-$ & $-$ & $-$\\
    M14 & $-$ & $-$ & Ns-M & Sl-M & $-$ & $-$\\
    M15 & $-$ & $-$ & $-$ & $-$ & Ns-M & Sl-M\\
    M16 & Ns-M & Sl-H & $-$ & $-$ & $-$ & $-$\\
    M17 & $-$ & $-$ & Ns-M & Sl-H & $-$ & $-$\\
    M18 & $-$ & $-$ & $-$ & $-$ & Ns-M & Sl-H\\
    M19 & Ns-H & Sl-L & $-$ & $-$ & $-$ & $-$\\
    M20 & $-$ & $-$ & Ns-H & Sl-L & $-$ & $-$\\
    M21 & $-$ & $-$ & $-$ & $-$ & Ns-H & Sl-L\\
    M22 & Ns-H & Sl-M & $-$ & $-$ & $-$ & $-$\\
    M23 & $-$ & $-$ & Ns-H & Sl-M & $-$ & $-$\\
    M24 & $-$ & $-$ & $-$ & $-$ & Ns-H & Sl-M\\
    M25 & Ns-H & Sl-H & $-$ & $-$ & $-$ & $-$\\
    M26 & $-$ & $-$ & Ns-H & Sl-H & $-$ & $-$\\
    M27 & $-$ & $-$ & $-$ & $-$ & Ns-H & Sl-H\\
    M28 & Ns-L & Sl-L & $-$ & $-$ & $-$ & $-$\\
    M29 & $-$ & $-$ & Ns-L & Sl-L & $-$ & $-$\\
    M30 & $-$ & $-$ & $-$ & $-$ & Ns-L & Sl-L\\
    M31 & Ns-M & Sl-M & $-$ & $-$ & $-$ & $-$\\
    M32 & $-$ & $-$ & Ns-M & Sl-M & $-$ & $-$\\
    M33 & $-$ & $-$ & $-$ & $-$ & Ns-M & Sl-M\\
    M34 & Ns-H & Sl-H & $-$ & $-$ & $-$ & $-$\\
    M35 & $-$ & $-$ & Ns-H & Sl-H & $-$ & $-$\\
    M36 & $-$ & $-$ & $-$ & $-$ & Ns-H & Sl-H\\
    M37 & Ns-L & Sl-L & Ns-M & Sl-M & Ns-H & Sl-H\\
    M38 & Ns-H & Sl-H & Ns-M & Sl-M & Ns-L & Sl-L\\
    M39 & Ns-M & Sl-M & Ns-L & Sl-L & $-$ & $-$\\
    M40 & Ns-M & Sl-M & Ns-H & Sl-H & $-$ & $-$\\
    M41 & $-$ & $-$ & Ns-M & Sl-M & Ns-L & Sl-L\\
    M42 & $-$ & $-$ & Ns-M & Sl-M & Ns-H & Sl-H\\
    M43 & Ns-L & Sl-L & Ns-L & Sl-L & Ns-L & Sl-L\\
    M44 & Ns-M & Sl-M & Ns-M & Sl-M & Ns-M & Sl-M\\
    M45 & Ns-H & Sl-H & Ns-H & Sl-H & Ns-H & Sl-H\\
    M46 & Ns-H & Sl-L & Ns-H & Sl-L & Ns-H & Sl-L\\
    M47 & Ns-L & Sl-H & Ns-L & Sl-H & Ns-L & Sl-H\\
    M48 & Ns-L & Sl-M & Ns-L & Sl-H & $-$ & $-$\\
    M49 & $-$ & $-$ & Ns-L & Sl-M & Ns-L & Sl-H\\
    M50 & $-$ & $-$ & Ns-H & Sl-M & Ns-H & Sl-M\\
    M51 & Ns-H & Sl-M & Ns-H & Sl-M & $-$ & $-$\\
    M52 & Ns-H & Sl-L & Ns-H & Sl-L & $-$ & $-$\\
    M53 & $-$ & $-$ & Ns-H & Sl-L & Ns-H & Sl-L\\
    \hline
    \end{tabular}
}
\end{table}
In general, audios can be characterized using the following parameters: sampling rate or number of samples per second, number of bits per sample also called bit depth (e.g. 8, 16 or 24 bits) and the number of channels (e.g. 1 channel for mono, 2 channels for stereo). The main characteristics of each original audio are presented in Table V.

\begin{table}[ht]
\caption{{CHARACTERISTICS OF ORIGINALS AUDIOS USED IN SUBJECTIVE TESTS}}
\label{tab:tab5}
\centering
\resizebox{\columnwidth}{!}{%
\begin{tabular}{>{\raggedright\arraybackslash}m{1.85cm}>{\centering\arraybackslash}m{1.75cm}
                >{\centering\arraybackslash}m{1.75cm}>{\centering\arraybackslash}m{1.75cm}}
    \hline
    \multicolumn{1}{c}{{\textbf{PARAMETER}}} & {\textbf{MUSIC}} & {\textbf{SPORT}} & {\textbf{NEWS}}\\
    \hline
    \hline
    {Codec and Sampling rate (kbps)} & AAC-LC - 576 {HE-AACv2-96} & AAC-LC - 576 {HE-AACv2-96} & AAC-LC - 576 {HE-AACv2-96}\\
    {Number of channels} & {2} & {2}& {2}\\
    {Number of bits per sample} & {16} & {16} & {16}\\
    {Audio length (seconds)} & {120} & {120} & {120}\\
    \hline
\end{tabular}
}
\end{table}

\vspace{\baselineskip}
   
Furthermore, in order to study the impact of the user preference on the global user’s QoE, another database was built; in which the explicit user preference for a content type is used; thus, each assessor manifests his or her preference before evaluate an audio file. In this database, only 10 impairment models from Table IV \textit{(M2, M19, M20, M21, M25, M26, M27, M43, M44 and M45)} were considered for each content type; and also two audio codecs were used. Then, a total number of 60 impairment audio files were generated. The audio characteristics are the same that those presented in Table V.

Finally, two extra audios were considered, the difference of these audios is their length, each one with a length of 20 minutes. The goal of these audio files is to test the processing and energy consumption of the proposed AsQM that is installed in a mobile handled device. For this test, the perception of users and their quality-of-experience is considered. Currently, one of the most important constraints on mobile devices is the energy consumption; therefore, the performance assessment of AsQM considering that aspect is relevant.
  
\section{Proposed Network Architecture}
\addcontentsline{toc}{section}{Proposed Network Architecture}

The scenario in which the \textit{AsQM} is implemented is described in this section.

\subsection{Customized Player in client side}
\addcontentsline{toc}{subsection}{Customized Player in client side}
   
A player was customized to monitor and capture parameters and states of the buffer, providing conditions to estimate the user QoE in the streaming audio service. 

The parameters captured from the buffer are: (a) period of the initial buffer; (b) period of playing, in which the audio is displayed continuously without interruptions; (c) period of rebuffering, during this period, temporal interruptions appear. Considering these parameters, the number, length and temporal location of stalls can be obtained. Also, the duration of the initial buffering delay can be determined and stored.
  
\subsection{Audio Database in server side}
\addcontentsline{toc}{subsection}{Audio Database in server side}
   
The $AsQM$ was determined through the results of the subjective tests. Several audios with different characteristics were considered in the tests. Thus, the audio categories considered were: music, sport and news.

The audio length considering in this work was the same of other studies \cite{9123118}, which uses audio tests and the MOS classification. As previously stated, a database of different audio files was considered with the following data: audio content type, audio length, number of stalls or pauses, the length of each pause and the specific time of occurrence (timestamp), and the user preference for the audio content (category). 
  
\subsection{Audio Service Application Scenario}
\addcontentsline{toc}{subsection}{Audio Service Application Scenario}
   
Subjective tests were performed in a laboratory environment, users were invited to listen to audios with different kinds of impairments and instructed to evaluate each audio quality based on the $ACR$ method which use the MOS scale.
  
\begin{figure}[ht]
	\begin{center}
		\includegraphics[width=3.27in,height=2.12in]{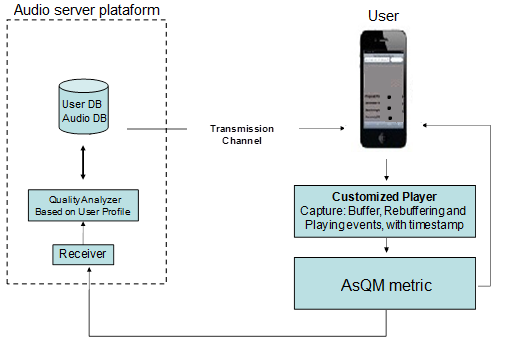}
		\caption{Proposed network architecture of audio streaming service implemented at the end-client.}
		\label{fig:fig7}
	\end{center}
\end{figure}

The $AsQM$ metric was implemented in a client mobile device used in the test. In a first phase, assessors went to a laboratory and an explanation was given, and each assessor fills a questionnaire with his or her user’s profile and scores each audio file. Fig. \ref{fig:fig7} presents the network architecture with emphasis on the client side implementation.

As second step, audio files and users’ profiles are stored in the server platform. The $AsQM$ score is transmitted from the client device and used with the audio identification ($V\_id$) and user’s preference as inputs in the algorithm that calculates the $AsQM$. This algorithm is presented in Table VI.
  
The application was built in 1 core processor of 1.6 GHz octa-core, the application for using the metric has a simple complexity.

\begin{table}[ht]
\caption{Algorithm of the proposed AsQM that considers users' Preference}
\label{tab:tab7}
\centering
\resizebox{\columnwidth}{!}{%
\begin{tabular}{>{\centering\arraybackslash}m{0.75cm}>{\raggedright\arraybackslash}m{6.95cm}}
    \hline
    \multicolumn{1}{c}{\textbf{LINE}} & \multicolumn{1}{c}{\textbf{STATEMENT}}\\
    \hline
    \hline
    1 & \textit{Audio ID}$: A\textsubscript{id}$\\
    2 & $CT\textsubscript{1}$: \textit{sport}$ =\{A\textsubscript{s1}, A\textsubscript{s2}, A\textsubscript{s3}, \cdots, A\textsubscript{si}\} // i.e.; A\textsubscript{id} = A\textsubscript{s1}$\\
    3 & $CT\textsubscript{2}:$ \textit{documentary} $\{A\textsubscript{d1}, A\textsubscript{d2}, A\textsubscript{d3}, \cdots, A\textsubscript{dj}\}$\\
    4 & $CT\textsubscript{3}:$ \textit{news} $\{A\textsubscript{n1}, A\textsubscript{n2}, A\textsubscript{n3}, \cdots, A\textsubscript{nk}\}$\\
    5 & $Up =$ \textit{User preference for audio CT}\\
    6 & $//$ \textit{Example}$:$ \textit{Up}$=\{ CT\textsubscript{1}, CT\textsubscript{2}\}$\\
    7 & \textit{Read} $(AsQM)$\\
    8 & \textit{If} $(A\textsubscript{id} \epsilon Up)$\\
    9 & $AsQM = AsQM\textsubscript{1} * PF_{p}^{CT}$\\
    10 & \textit{Else}\\
    11 & $AsQM = AsQM\textsubscript{1} * PF_{np}^{CT}$\\
    \hline
\end{tabular}
}
\end{table}
   
All audios had a duration of 60 seconds and were divided in 3 segments: (1) 0 seconds to 19 seconds, (2) 20 seconds to 39 seconds and (3) 40 seconds to 60 seconds. As presented in Table III, three different impairment groups based on stall characteristics were defined. 

As stated before, it is also a goal of this research to find out if there are any differences in the user perceived quality when considering different audio categories, therefore, three content categories were used as test material.
  
\section{RESULTS}
\addcontentsline{toc}{section}{RESULTS}
   
In the phase A of this study, subjective tests were performed in a laboratory environment to study the influence of pauses location on users' QoE about audio files. In the phase B, the performance of the $AsQM$ was validated considering (i) correlation with subjective MOS scores that includes the user preference on audio content, and (ii) impact on the processing and energy consumption on current mobile hand-held devices.
  
\subsection{Phase A: study of influence of temporal location of pauses and initial delay on users' QoE}
\addcontentsline{toc}{subsection}{Phase A: study of user QoE influence of temporal location of pauses in audio files}
   
The total number of assessors used in the tests was 96, and each audio had\ at least fifteen scores.  All assessors reported to have no hearing impairment and have no experience in assessing audio quality tests. 

\begin{figure}[htb!]
	\begin{center}
		\includegraphics[width=2.93in,height=2.39in]{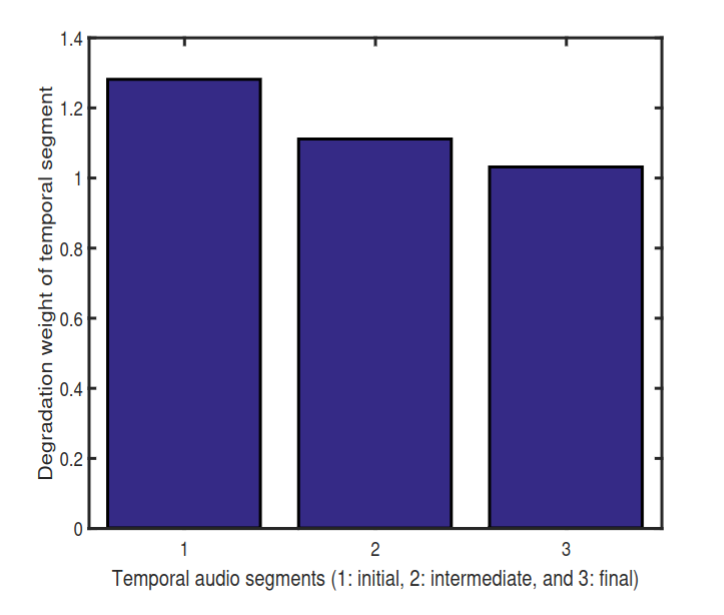}
		\caption{Degradation weights of initial (1), intermediate (2) and final (3) audio temporal segment, denoted by $D_A$, $D_B$ and $D_C$, respectively.}
		\label{fig:fig9}
	\end{center}
\end{figure}

As presented in Table IV, stalls in different location and several audio length for each audio category were considered. The average MOS value of the three audio categories for the same scenario was considered. Thus, using (7) for each impairment model and considering the two audio codecs, we built the equation linear system defined in (9). Thus, the values of C and the degradation weights of temporal segments $D_A$, $D_B$, and $D_C$ were obtained. These weights values are presented in Fig. \ref{fig:fig9}.

Fig. \ref{fig:fig9} shows that the initial audio segment is more relevant, or it has more impairment weight, in relation to other temporal segments. The pauses at the beginning of the audio have more negative effect on user QoE.

The assessors reported that if there are disturbances at the beginning of the audio, they became pessimistic and think the pauses could happen throughout all the audio.

Regarding the subjective test results of initial buffering delay, it was possible to determine the variables introduced in (6): k=0.824 and c=1.017. For this experimental tests, we used audio lengths of 30, 60 and 90 seconds with initial delay lengths of 2, 4, 6, 8, 10, 14, 18 and 22 seconds. Considering the results obtained from subjective tests and the objective measure using (6), a  Pearson Correlation Coefficient (PCC) of 0.974 and a maximum error of 0.265 at 5-point MOS scale were calculated.
  
\subsection{Phase B: study of the impact of audio content preference on users' QoE}
\addcontentsline{toc}{subsection}{Phase B: study of user preference influence in QoE of audio files}
   
In a first phase, subjective tests were performed in a laboratory environment using the audio application implemented in a specific mobile device, and the results determined the $PF_{p}^{CT}$ and $PF_{np}^{CT}$functions, and later, the $AsQM$ metric was determined. In an extended test, the performance of the $AsQM$ was evaluated using the crowdsourcing method. Both, face and remote tests are explained as follows.

A sound card with technologies including stunning 3D surround effects was used in the face--to--face tests,as well as a headphone of 3.5mm jack input and stereo sound with 5W of power. The room environment had no disturbing objects or external noise. The tests were conducted by a period of 6.5 weeks. The tests were performed individually without a time limit, in which instructions was performed about the tests, before the valid tests. 


The functions $PF_{p}^{CT}$ and $PF_{np}^{CT}$ depend on both the MOS scores and the audio categories. According to the $PF_{p}^{CT}$, the sport content presents the lowest values, and news content obtained the highest values. Thus, the users with preference for sport content are more negatively affected.

   
In addition, assessors evaluated the perceived consumed resources using a five-point MOS scale. The experimental results are presented in Table VII.

\begin{table}[ht]
\caption{Perceived Value of Consumed Resources in the Mobile Device}
\label{tab:tab9}
\centering
\resizebox{\columnwidth}{!}{%
\begin{tabular}{>{\raggedright\arraybackslash}m{4.60cm}>{\centering\arraybackslash}m{3.10cm}}
    \hline
    \multicolumn{1}{c}{\textbf{PARAMETER}} & \textbf{AVERAGE VALUE (1-5)}\\
    \hline
    \hline
    Processing and memory consumption & 4.5\\
    Energy consumption & 4.5\\
    \hline
\end{tabular}
}
\end{table}
   
These results demonstrated that there was not a negative impact on the performance of the mobile device used in the face tests. This is because the proposed metric is based only on the application layer parameters and the metric added to the audio player consumes very low processing. For users analyze whether the audio player was consuming resources, was initially presented the player without and with the addition of proposal metric.
  
In the extended tests the performance evaluation of the $AsQM$ was performed using the remote method, crowdsourcing, in which the assessors had different audio card, headphone or loudspeaker and mobile devices linked to network, the instruction was that the user had to download the audio to hear after the audio by the audio player. Additional 12 audios were used as test material, in which 4 audios for each category were considered and a different number of pauses were inserted. In these tests, audio of 60 seconds are used. 
 
The crowdsourcing platform established that as minimum 30 users need to participate in each campaign. Thus, three campaigns were sent for each audio, achieving 90 MOS scores for each audio listened. The results of both remote and face-to-face tests are shown in Table VIII. 

\begin{table}[ht]
\caption{Audio Quality Tests Results Performed in a Laboratory Environment and Using Crowdsourcing Method}
\label{tab:tab10}
\centering
\resizebox{\columnwidth}{!}{%
\begin{tabular}{>{\centering\arraybackslash}m{1.60cm}>{\centering\arraybackslash}m{1.60cm}
                >{\centering\arraybackslash}m{1.90cm}>{\centering\arraybackslash}m{2.45cm}}
    \hline
    \textbf{AUDIO CATEGORY} & \textbf{AUDIO SEQUENCE} & \textbf{LABORATORY} & \textbf{CROWDSOURCING}\\
    \hline
    \hline
    Sport & 1 & {3,25} &  {3,28}\\
    Sport & 2 &  {2.87} &  {2,94}\\
    Documentary & 3 &  {3,52} &  {3,63}\\
    Documentary & 4 &  {2,98} &  {2,87}\\
    News & 5 &  {3,72} &  {3,79}\\
    News & 6 &  {3,22} &  {3,34}\\
    \hline
\end{tabular}
}
\end{table}

As can be observed in Table VIII the results of both subjective test methodologies obtained similar scores.
   
\begin{figure}[ht]
	\begin{center}
		\includegraphics[width=3.63in,height=2.1in]{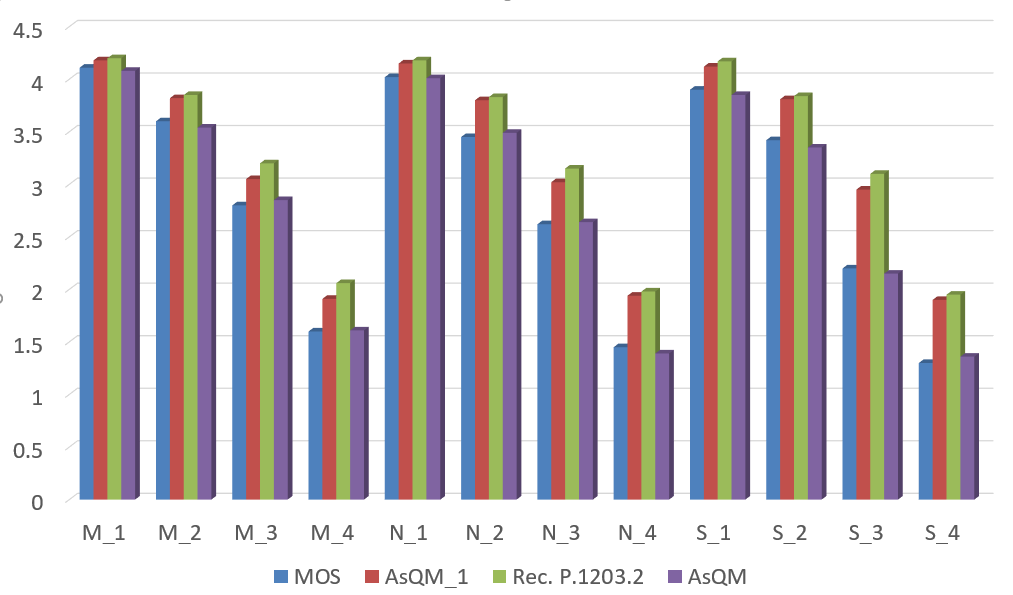}
		\caption{Performance assessment comparison of the objective MOS determined by $AsQM_1$ and $AsQM$.}
		\label{fig:fig11}
	\end{center}
\end{figure}

Fig. \ref{fig:fig11} presents the MOS values obtained by crowdsourcing method, ITU-T recommendation, $AsQM_1$ and $AsQM$, in which the assessors evaluated the audio content and previously he/she manifested his/her preference. As previously stated, audio content about sport, music and news were used. Audio sequences indexed named $``$1$"$ to $``$4$"$  represent the audios with the lowest and highest number of pauses, respectively. Audios indexed as 2 and 3 correspond to the medium impairment group described in Table III. 

It is observed that sport audios obtained the lowest MOS indexes. On the other hand, music content presented a lower impact in relation to the other contents, maybe because some users know the music and lyric, and it masked the negative impact, but further studies are necessary to understand this user behavior. In general, all objective metrics obtained satisfactory results in the scenarios with low degradation. The metrics that not considered the user subjectivity (preference factor), $AsQM_1$ and from ITU, obtained almost the same scores, but slightly better correlation with MOS scores was reached by the $AsQM_1$.
A Pearson correlation coefficient and a Root mean square error of 0.99 and 0.23, respectively, were calculated between AsQM scores and subjective test scores.
  
\section{CONCLUSION}
\addcontentsline{toc}{section}{CONCLUSIONS}
   
The aim of broadcast technology is to deliver information and entertainment to audiences worldwide, then, the quality of the information that people receive is relevant. In this arena, the present research developed a metric specifically designed to address the issue of audio transmission over IP networks, but also including the subjectivity of the users. Thus, different audio content is considered in the database used as test material. Also, the study aimed to understand the negative impacts caused by stalls in different segments of the transmission, as well as find out if the occurrence of stalls results in different quality perceptions.

This paper stressed the reasons why current subjective test methodologies are not correlated with real services with the user QoE about audio streaming service. The subjective tests results showed the relevance of considering the temporal location of pauses in an audio quality assessment model.
 
The results highlighted the relevance of considering the user preference for audio content, in which assessors with preference granted the lowest MOS scores for the presented audios. Thus, the performance of diverse objective metrics can be enhanced.
Functions $PF_{p}^{CT}$ and $PF_{np}^{CT}$ were defined, which work as a correction factor and are meant to consider the user subjectivity into an objective metric. The experimental results demonstrated that these functions are able to enhance the performance of existing audio quality metrics.

Additionally, the metric was evaluated according to consuming resources and the results show the inclusion of the mathematical model solution in a sound player consumes very low resources from current electronic devices. The assessors evaluated an imperceptible interruption regarding the consumed resources in the electronic device.

\bibliographystyle{ieeetr}
\bibliography{refbib}

\begin{thebibliography}{10}

\bibitem{JAALAMA2021103996}
K.~Jaalama, N.~Fagerholm, A.~Julin, J.-P. Virtanen, M.~Maksimainen, and
  H.~Hyyppä, ``Sense of presence and sense of place in perceiving a 3d
  geovisualization for communication in urban planning – differences
  introduced by prior familiarity with the place,'' {\em Landscape and Urban
  Planning}, vol.~207, p.~103996, 2021.

\bibitem{9022990}
G.~Li and R.~Liu, ``Original music album diffusion sustainability in social
  network-based communities: A network embedded perspective,'' {\em IEEE
  Access}, vol.~8, pp.~53107--53115, 2020.

\bibitem{9777989}
J.~Skowronek, A.~Raake, G.~H. Berndtsson, O.~S. Rummukainen, P.~Usai, S.~N.~B.
  Gunkel, M.~Johanson, E.~A.~P. Habets, L.~Malfait, D.~Lindero, and A.~Toet,
  ``Quality of experience in telemeetings and videoconferencing: A
  comprehensive survey,'' {\em IEEE Access}, vol.~10, pp.~63885--63931, 2022.

\bibitem{3210424.3210431}
A.~Trattnig, C.~Timmerer, and C.~Mueller, ``Investigation of youtube regarding
  content provisioning for http adaptive streaming,'' in {\em Proceedings of
  the 23rd Packet Video Workshop}, PV '18, (New York, NY, USA), p.~60–65,
  Association for Computing Machinery, 2018.

\bibitem{ishizaki2015understanding}
H.~Ishizaki, S.~C. Herring, G.~Hattori, and Y.~Takishima, ``Understanding user
  behavior on online music distribution sites: A discourse approach,'' {\em
  iConference 2015 Proceedings}, 2015.

\bibitem{9166225}
R.~A. Rahimi and K.-H. Park, ``A comparative study of internet architecture and
  applications of online music streaming services: The impact on the global
  music industry growth,'' in {\em 2020 8th International Conference on
  Information and Communication Technology (ICoICT)}, pp.~1--6, 2020.

\bibitem{SEYFOLLAHI2020107368}
A.~Seyfollahi and A.~Ghaffari, ``A lightweight load balancing and route
  minimizing solution for routing protocol for low-power and lossy networks,''
  {\em Computer Networks}, vol.~179, p.~107368, 2020.

\bibitem{protocol1980rfc}
U.~D. Protocol, ``Rfc 768 j. postel isi 28 august 1980,'' {\em Isi}, 1980.

\bibitem{rix2001perceptual}
A.~W. Rix, J.~G. Beerends, M.~P. Hollier, and A.~P. Hekstra, ``Perceptual
  evaluation of speech quality (pesq)-a new method for speech quality
  assessment of telephone networks and codecs,'' in {\em 2001 IEEE
  International Conference on Acoustics, Speech, and Signal Processing.
  Proceedings (Cat. No. 01CH37221)}, vol.~2, pp.~749--752, IEEE, 2001.

\bibitem{chaudhery2017performance}
T.~J. Chaudhery, ``Performance evaluation of codel queue mechanism and tfrc
  transport protocol when using voip flows,'' in {\em 2017 International
  Conference on Frontiers of Information Technology (FIT)}, pp.~1--6, IEEE,
  2017.

\bibitem{duanmu2018quality}
Z.~Duanmu, A.~Rehman, and Z.~Wang, ``A quality-of-experience database for
  adaptive video streaming,'' {\em IEEE Transactions on Broadcasting}, vol.~64,
  no.~2, pp.~474--487, 2018.

\bibitem{8743302}
A.~Ragano, E.~Benetos, and A.~Hines, ``Adapting the quality of experience
  framework for audio archive evaluation,'' in {\em 2019 Eleventh International
  Conference on Quality of Multimedia Experience (QoMEX)}, pp.~1--3, 2019.

\bibitem{8657699}
D.~Z. Rodríguez, R.~L. Rosa, F.~L. Almeida, G.~Mittag, and S.~Möller,
  ``Speech quality assessment in wireless communications with mimo systems
  using a parametric model,'' {\em IEEE Access}, vol.~7, pp.~35719--35730,
  2019.

\bibitem{8926509}
J.~G. Beerends, N.~M.~P. Neumann, E.~L. van~den Broek, A.~Llagostera~Casanovas,
  J.~T. Menendez, C.~Schmidmer, and J.~Berger, ``Subjective and objective
  assessment of full bandwidth speech quality,'' {\em IEEE/ACM Transactions on
  Audio, Speech, and Language Processing}, vol.~28, pp.~440--449, 2020.

\bibitem{9281368}
R.~Hassen, B.~Gülecyüz, and E.~Steinbach, ``Pvc-slp: Perceptual
  vibrotactile-signal compression based-on sparse linear prediction,'' {\em
  IEEE Transactions on Multimedia}, vol.~23, pp.~4455--4468, 2021.

\bibitem{7797329}
E.~T. Affonso, D.~Z. Rodríguez, R.~L. Rosa, T.~Andrade, and G.~Bressan,
  ``Voice quality assessment in mobile devices considering different fading
  models,'' in {\em 2016 IEEE International Symposium on Consumer Electronics
  (ISCE)}, pp.~21--22, 2016.

\bibitem{9351640}
D.~Z. Rodríguez, D.~Carrillo, M.~A. Ramírez, P.~H.~J. Nardelli, and
  S.~Möller, ``Incorporating wireless communication parameters into the
  e-model algorithm,'' {\em IEEE/ACM Transactions on Audio, Speech, and
  Language Processing}, vol.~29, pp.~956--968, 2021.

\bibitem{Rao2019}
K.~S. Rao and N.~P. Narendra, {\em Hybrid Approach of Modeling the Source
  Signal}, pp.~75--103.
\newblock Cham: Springer International Publishing, 2019.

\bibitem{8937202}
H.~Gamper, C.~K.~A. Reddy, R.~Cutler, I.~J. Tashev, and J.~Gehrke, ``Intrusive
  and non-intrusive perceptual speech quality assessment using a convolutional
  neural network,'' in {\em 2019 IEEE Workshop on Applications of Signal
  Processing to Audio and Acoustics (WASPAA)}, pp.~85--89, 2019.

\bibitem{GOOSSENS2017109}
T.~Goossens, C.~Vercammen, J.~Wouters, and A.~{van Wieringen}, ``Masked speech
  perception across the adult lifespan: Impact of age and hearing impairment,''
  {\em Hearing Research}, vol.~344, pp.~109--124, 2017.

\bibitem{SONI202127}
M.~H. Soni and H.~A. Patil, ``Non-intrusive quality assessment of
  noise-suppressed speech using unsupervised deep features,'' {\em Speech
  Communication}, vol.~130, pp.~27--44, 2021.

\bibitem{app10051793}
L.~Du, L.~Zhuo, J.~Li, J.~Zhang, X.~Li, and H.~Zhang, ``Video quality of
  experience metric for dynamic adaptive streaming services using dash standard
  and deep spatial-temporal representation of video,'' {\em Applied Sciences},
  vol.~10, no.~5, 2020.

\bibitem{rodriguez2014impact}
D.~Z. Rodr{\'\i}guez, Z.~Wang, R.~L. Rosa, and G.~Bressan, ``The impact of
  video-quality-level switching on user quality of experience in dynamic
  adaptive streaming over http,'' {\em EURASIP Journal on Wireless
  Communications and Networking}, vol.~2014, no.~1, pp.~1--15, 2014.

\bibitem{9388867}
M.~Torcoli, T.~Kastner, and J.~Herre, ``Objective measures of perceptual audio
  quality reviewed: An evaluation of their application domain dependence,''
  {\em IEEE/ACM Transactions on Audio, Speech, and Language Processing},
  vol.~29, pp.~1530--1541, 2021.

\bibitem{9075375}
X.~Min, G.~Zhai, J.~Zhou, M.~C.~Q. Farias, and A.~C. Bovik, ``Study of
  subjective and objective quality assessment of audio-visual signals,'' {\em
  IEEE Transactions on Image Processing}, vol.~29, pp.~6054--6068, 2020.

\bibitem{rodriguez2016video}
D.~Z. Rodr{\'\i}guez, R.~L. Rosa, E.~C. Alfaia, J.~I. Abrah{\~a}o, and
  G.~Bressan, ``Video quality metric for streaming service using dash
  standard,'' {\em IEEE Transactions on broadcasting}, vol.~62, no.~3,
  pp.~628--639, 2016.

\bibitem{affonso2018speech}
E.~T. Affonso, R.~D. Nunes, R.~L. Rosa, G.~F. Pivaro, and D.~Z. Rodriguez,
  ``Speech quality assessment in wireless voip communication using deep belief
  network,'' {\em IEEE Access}, vol.~6, pp.~77022--77032, 2018.

\bibitem{rodriguez2012quality}
D.~Z. Rodriguez, J.~Abrahao, D.~C. Begazo, R.~L. Rosa, and G.~Bressan,
  ``Quality metric to assess video streaming service over tcp considering
  temporal location of pauses,'' {\em IEEE Transactions on Consumer
  Electronics}, vol.~58, no.~3, pp.~985--992, 2012.

\bibitem{duanmu2016quality}
Z.~Duanmu, K.~Zeng, K.~Ma, A.~Rehman, and Z.~Wang, ``A quality-of-experience
  index for streaming video,'' {\em IEEE Journal of Selected Topics in Signal
  Processing}, vol.~11, no.~1, pp.~154--166, 2016.

\bibitem{bampis2017study}
C.~G. Bampis, Z.~Li, A.~K. Moorthy, I.~Katsavounidis, A.~Aaron, and A.~C.
  Bovik, ``Study of temporal effects on subjective video quality of
  experience,'' {\em IEEE Transactions on Image Processing}, vol.~26, no.~11,
  pp.~5217--5231, 2017.

\bibitem{robitza2018http}
W.~Robitza, S.~G{\"o}ring, A.~Raake, D.~Lindegren, G.~Heikkil{\"a},
  J.~Gustafsson, P.~List, B.~Feiten, U.~W{\"u}stenhagen, M.-N. Garcia, {\em
  et~al.}, ``Http adaptive streaming qoe estimation with itu-t rec. p. 1203:
  open databases and software,'' in {\em Proceedings of the 9th ACM Multimedia
  Systems Conference}, pp.~466--471, 2018.

\bibitem{8436879}
T.~He, R.~Xie, J.~Su, X.~Tang, and L.~Song, ``A no reference bitstream-based
  video quality assessment model for h.265/hevc and h.264/avc,'' in {\em 2018
  IEEE International Symposium on Broadband Multimedia Systems and Broadcasting
  (BMSB)}, pp.~1--5, 2018.

\bibitem{9234526}
A.~Raake, S.~Borer, S.~M. Satti, J.~Gustafsson, R.~R.~R. Rao, S.~Medagli,
  P.~List, S.~Göring, D.~Lindero, W.~Robitza, G.~Heikkilä, S.~Broom,
  C.~Schmidmer, B.~Feiten, U.~Wüstenhagen, T.~Wittmann, M.~Obermann, and
  R.~Bitto, ``Multi-model standard for bitstream-, pixel-based and hybrid video
  quality assessment of uhd/4k: Itu-t p.1204,'' {\em IEEE Access}, vol.~8,
  pp.~193020--193049, 2020.

\bibitem{7965631}
A.~Raake, M.-N. Garcia, W.~Robitza, P.~List, S.~Göring, and B.~Feiten, ``A
  bitstream-based, scalable video-quality model for http adaptive streaming:
  Itu-t p.1203.1,'' in {\em 2017 Ninth International Conference on Quality of
  Multimedia Experience (QoMEX)}, pp.~1--6, 2017.

\bibitem{guzman2019automatic}
P.~Guzm{\'a}n~Castillo, P.~Arce~Vila, and J.~C. Guerri~Cebollada, ``Automatic
  qoe evaluation of dash streaming using itu-t standard p. 1203 and google
  puppeteer,'' in {\em Proceedings of the 16th ACM International Symposium on
  Performance Evaluation of Wireless Ad Hoc, Sensor, \& Ubiquitous Networks},
  pp.~79--86, 2019.

\bibitem{8475048}
E.~Liri, P.~K. Singh, A.~B. Rabiah, K.~Kar, K.~Makhijani, and K.~Ramakrishnan,
  ``Robustness of iot application protocols to network impairments,'' in {\em
  2018 IEEE International Symposium on Local and Metropolitan Area Networks
  (LANMAN)}, pp.~97--103, 2018.

\bibitem{itu1999subjective}
P.~ITU-T~RECOMMENDATION, ``Subjective video quality assessment methods for
  multimedia applications,'' {\em International telecommunication union}, 1999.

\bibitem{jaramillo2016content}
B.~O. Jaramillo~Sr, J.~O. Ni{\~n}o-Casta{\~n}eda, L.~Plati{\v{s}}a, and
  W.~Philips, ``Content-aware objective video quality assessment,'' {\em
  Journal of Electronic Imaging}, vol.~25, no.~1, p.~013011, 2016.

\bibitem{vandenbroucke2015subjective}
K.~Vandenbroucke, E.~P.~N. Castellar, and L.~De~Marez, ``Subjective insights
  from time and place shifters in assessing temporal quality of experience,''
  in {\em 2015 Seventh International Workshop on Quality of Multimedia
  Experience (QoMEX)}, pp.~1--2, IEEE, 2015.

\bibitem{laiche2021machine}
F.~Laiche, A.~B. Letaifa, I.~Elloumi, and T.~Aguili, ``When machine learning
  algorithms meet user engagement parameters to predict video qoe,'' {\em
  Wireless Personal Communications}, vol.~116, no.~3, pp.~2723--2741, 2021.

\bibitem{rosa2013sentimeter}
R.~L. Rosa, D.~Z. Rodriguez, and G.~Bressan, ``Sentimeter-br: A social web
  analysis tool to discover consumers' sentiment,'' in {\em 2013 IEEE 14th
  International Conference on Mobile Data Management}, vol.~2, pp.~122--124,
  IEEE, 2013.

\bibitem{rosa2013sentimeter2}
R.~L. Rosa, D.~Z. Rodriguez, and G.~Bressan, ``Sentimeter-br: A new social web
  analysis metric to discover consumers' sentiment,'' in {\em 2013 IEEE
  International Symposium on Consumer Electronics (ISCE)}, pp.~153--154, IEEE,
  2013.

\bibitem{rodriguez2014video}
D.~Z. Rodr{\'\i}guez, R.~L. Rosa, E.~A. Costa, J.~Abrah{\~a}o, and G.~Bressan,
  ``Video quality assessment in video streaming services considering user
  preference for video content,'' {\em IEEE Transactions on Consumer
  Electronics}, vol.~60, no.~3, pp.~436--444, 2014.

\bibitem{garcia2013parametric}
M.-N. Garcia, P.~List, S.~Argyropoulos, D.~Lindegren, M.~Pettersson, B.~Feiten,
  J.~Gustafsson, and A.~Raake, ``Parametric model for audiovisual quality
  assessment in iptv: Itu-t rec. p. 1201.2,'' in {\em 2013 IEEE 15th
  International Workshop on Multimedia Signal Processing (MMSP)}, pp.~482--487,
  IEEE, 2013.

\bibitem{9123118}
A.~Schwind, C.~Moldovan, T.~Janiak, N.~D. Dworschak, and T.~Hoßfeld, ``Don't
  stop the music: Crowdsourced qoe assessment of music streaming with
  stalling,'' in {\em 2020 Twelfth International Conference on Quality of
  Multimedia Experience (QoMEX)}, pp.~1--6, 2020.

\end{thebibliography}

\begin{IEEEbiography}[{\includegraphics[width=1in,height=1.25in,clip,keepaspectratio]{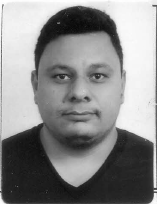}}]{Marcelo Rodrigo dos Santos}
Master's student at the Federal University of Lavras, (UFLA), Graduated in Computing from the Federal Institute of Education, Science and Technology of the South of Minas Gerais (IFSULDEMINAS - 2018 Campus Machado), degree in Technologist in Computer Networks from Higher Education Center e Research of Machado (CESEP - 2010) and specialization in Technologies for Web Development by the Federal Institute of Education, Science and Technology of the South of Minas Gerais (IFSULDEMINAS - 2020 Campus Passos). He is currently Network and Computer Systems ADM.  He has more than 10 years of experience in the field of Computing. 
His current research interests include computer networks, telecommunications systems, Natural Language Processing (NLP), quality of experience on multimedia services and data analysis from social networks.
\end{IEEEbiography}

\begin{IEEEbiography}[{\includegraphics[width=1in,height=1.25in,clip,keepaspectratio]{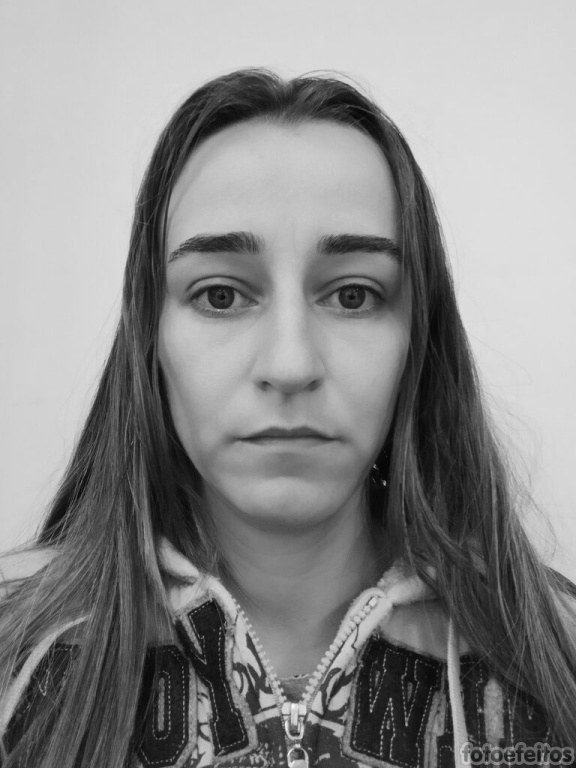}}]{Andreza Patrícia Batista}
 received a bachelor's degree in Electrical Engineering at the Federal Institute of Education, Science and Technology of Minas Gerais-Campus Formiga, Minas Gerais, Brazil. Currently studying for a master's degree in Systems Engineering and Automation from the Federal University of Lavras, Brasil. Also part of the effective staff, Electronics laboratory Technician of the Laboratory Sector, of the Federal Institute of Education, Science and Technology of Minas Gerais-Campus Formiga. His research interests include radiofrequency signal attenuation by atmospheric phenomena, 5G and 6G wireless networks, studies of long-term climatic magnitudes applied to different geographic coordinates and network simulator.
\end{IEEEbiography}

\begin{IEEEbiography}[{\includegraphics[width=1in,height=1.25in,clip,keepaspectratio]{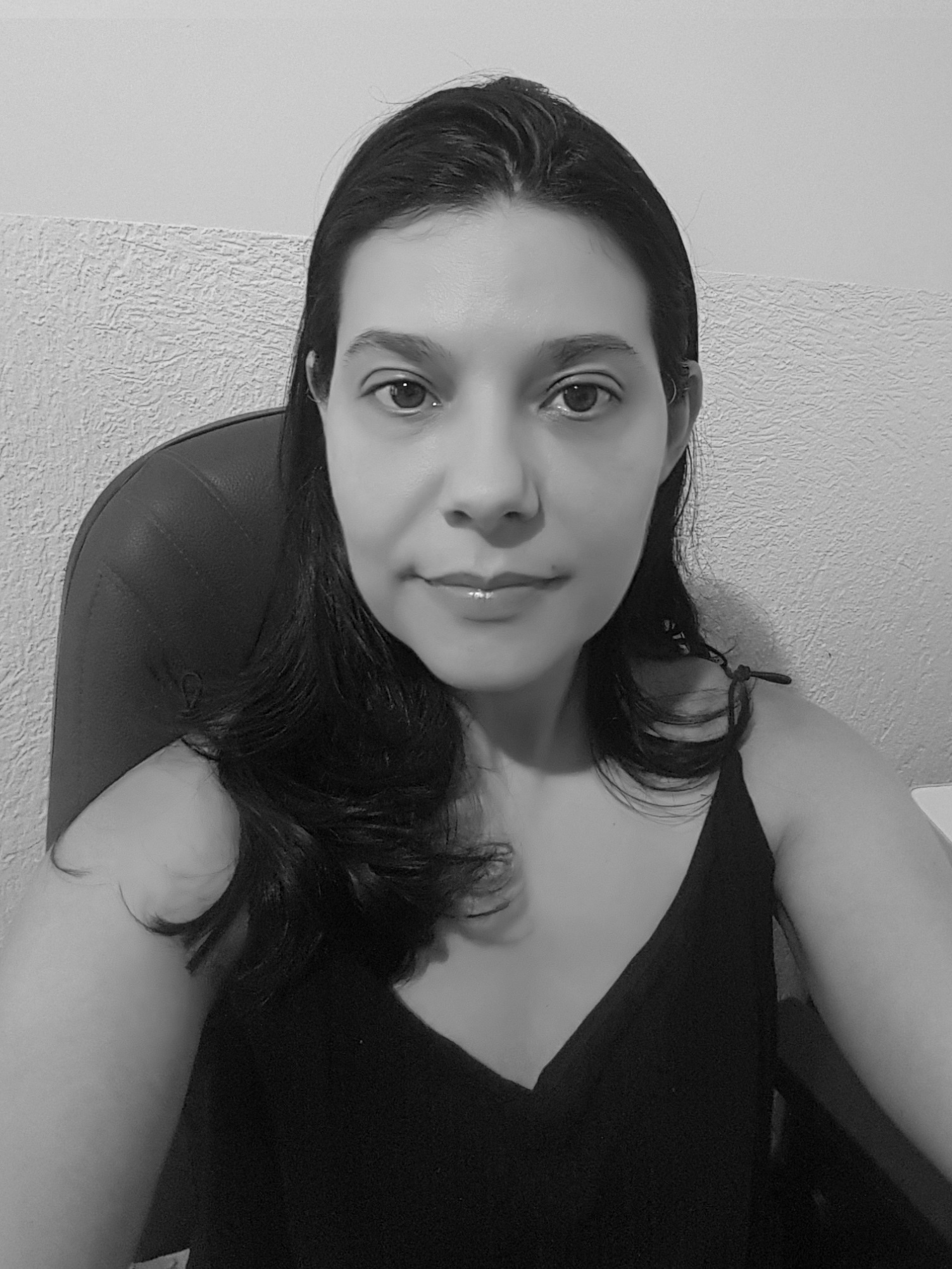}}]{Renata Lopes Rosa} received the M.S. degree from the University of São Paulo in 2009 and the Ph.D.
degree from the Polytechnic School of the University of São Paulo in 2015 (EPUSP). She is currently an Adjunct Professor with the Department of Computer Science, Federal University of Lavras, Brazil. She has a solid knowledge in computer science based on more than ten years of professional experience.
Her current research interests include computer networks, telecommunication systems, machine learning, quality of experience of multimedia service, cybersecurity, social networks, and recommendation systems.
\end{IEEEbiography}

\begin{IEEEbiography}[{\includegraphics[width=1in,height=1.25in,clip,keepaspectratio]{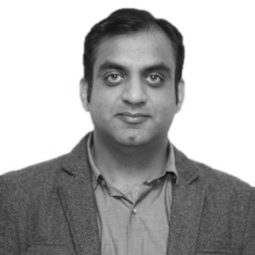}}]{Muhammad Saadi} (F’16) received a Ph.D. degree in electrical engineering from Chulalongkorn University, Bangkok, Thailand. He received his M.Sc and B.Sc degree from the National University of Malaysia, Malaysia and National University of Computer and Emerging Sciences, Pakistan in years 2009 and 2007 respectively. He was with National Electronics and Computer Technology Center, Aimagin Ltd., Thailand. He is currently an Associate Professor with the Department of Electrical Engineering, University of Central Punjab, Pakistan. His current research interests include visible light communication, indoor localization, machine learning, deep learning, and next-generation networks. 
\end{IEEEbiography}

\begin{IEEEbiography}[{\includegraphics[width=1in,height=1.25in,clip,keepaspectratio]{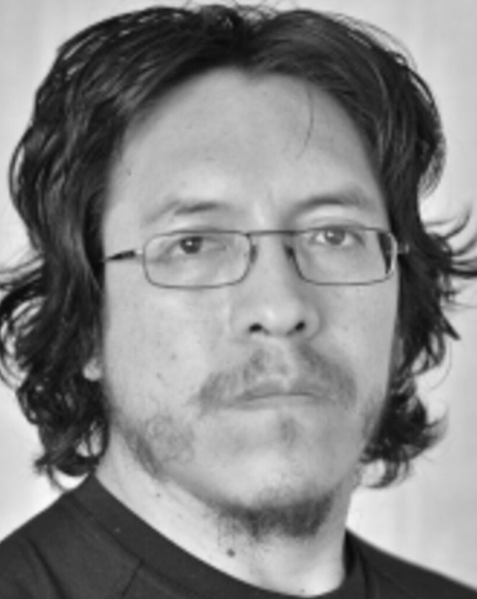}}]{Dick Carrillo} (Member, IEEE) 
received the B.Eng. degree (Hons.) in electronics
and electrical engineering from the National University of San Marcos, Lima, Peru, in 2004, and
the M.Sc. degree in electrical engineering from the
Pontifical Catholic University of Rio de Janeiro,
Rio de Janeiro, Brazil, in 2008. He is currently
pursuing the Ph.D. degrees in electrical engineering with the Lappeenranta–Lahti University of
Technology. From 2008 to 2010, he contributed to
WIMAX (IEEE 802.16m) Standardization. From 2010 to 2018, he worked
with the design and implementation of cognitive radio networks and projects
based on 3GPP technologies. Since 2018, he has been a Researcher at the
Lappeenranta–Lahti University of Technology. Since 2022, he has been a
Senior Standardization Specialist at Nokia Bell Labs, where he is contributing on shaping the 3GPP release 18 standard (5G-Advanced). His research
interests include mobile technologies beyond 5G, energy harvesting, intelligent meta-surfaces, cell-free mMIMO, and RAN slicing.
\end{IEEEbiography}

\begin{IEEEbiography}[{\includegraphics[width=1in,height=1.25in,clip,keepaspectratio]{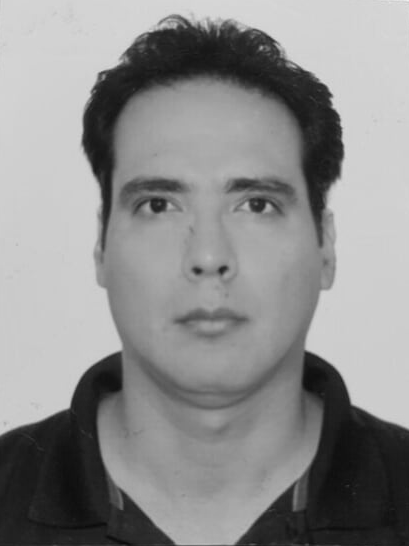}}]{Dem\'ostenes Zegarra Rodr\'iguez} (M'12-SM'15) received the B.S. degree in electronic engineering from the Pontifical Catholic University of Peru, the M.S. degree and the Ph.D. degree from the University of S\~ao Paulo in 2009 and 2013, respectively. In 2018-2019, he had a post-doctoral position in the Technical University of Berlin, specifically in the Quality and Usability laboratory,  He is currently an Adjunct Professor with the Department of Computer Science, Federal University of Lavras, Brazil. He has a solid knowledge in Telecommunication Systems and Computer Science based on 15 years of professional experience in major companies. His research interests include QoS and QoE in multimedia services, architect solutions in telecommunication systems, intrusion detection system, cybersecurity. He is a member of the Brazilian Telecommunications Society.
\end{IEEEbiography}


\end{document}